\documentclass[11pt]{article}

\usepackage{graphicx}

\begin{document}

\begin{flushright}
%New Foundations of Physics Institute preprint number
NFPI-10-1
\end{flushright}
\vspace{0.2 cm}

\begin{center}
%Title of paper
\Large{\bf The Michelson-Morley experiment in an accelerated reference frame}

\vspace{0.8 cm}
\renewcommand{\thefootnote}{\fnsymbol{footnote}}
\large{Dennis Crossley}\footnote{dennis.crossley@uwc.edu}

%\vspace{0.1 cm}
{\emph {Dept.~of Physics, University of Wisconsin-Sheboygan, Sheboygan, WI 53081}}

%\vspace{0.8 cm}
%\today
\end{center}

%\vspace{0.2 cm}
\begin{abstract}
We analyze the Michelson-Morley experiment in a reference frame moving with constant proper acceleration.   Interestingly, we find an expected fringe shift which depends not only on the interferometer's rate of acceleration, but also on its
speed relative to a preferred absolute reference frame.  While it has been repeatedly shown that no experiment performed in an inertial reference frame can detect that frame's absolute speed, the analysis in this paper suggests that by considering experiments in accelerated reference frames it may be possible to measure absolute speed after all.
\end{abstract}

\noindent
{\emph{Keywords: special relativity, Michelson-Morley experiment, accelerated reference frame, absolute space
\\PACS: 03.30.+p}} % Special Relativity
\vspace{0.2 cm}

\section{Introduction\label{intro}}
The Michelson-Morley experiment\cite{MMexpt} is one of the classic
experiments in the history of physics.  Its original purpose was to
measure the speed of the earth through the luminiferous ether by
measuring the shift of interference fringes in an interferometer.
The ether was, at the time, believed to be the medium through which
light waves traveled at speed $c$.  It was assumed that the earth
was moving at some speed $v$ relative to the lumiferous ether.  The
only property of the ether that concerns us here is that it is a
unique preferred reference frame in which light  travels at the same
speed in all directions.  In our analysis we will make reference to
this absolute reference frame and leave all references to the
luminiferous ether to the past.

\begin{figure}
\includegraphics[scale=.5]{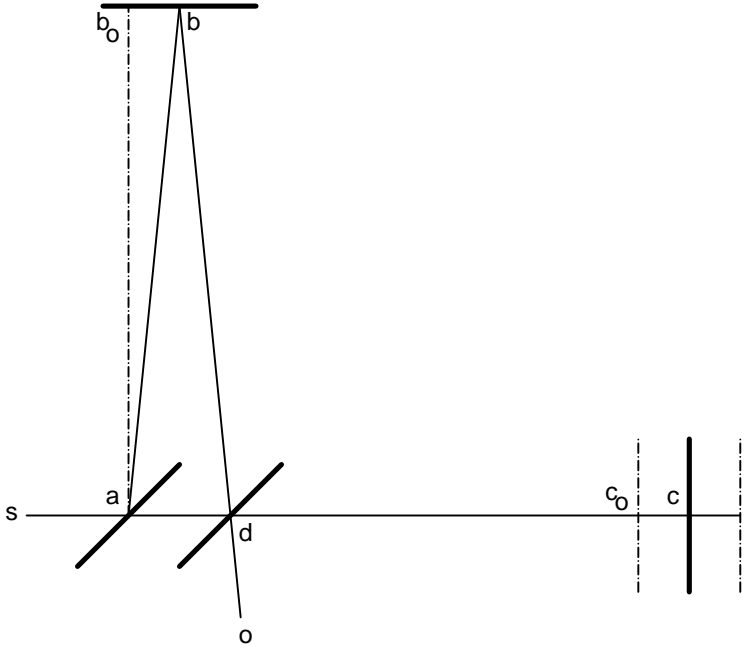}%
\caption{\label{fig:interferometer}Light paths in the
Michelson-Morley interferometer.}
\end{figure}

Fig.~\ref{fig:interferometer} shows the light paths through the
interferometer.
The experiment involves sending a beam of light from
a source $s$, splitting the beam at $a$ so that it follows two
perpendicular paths $abd$ and $acd$, recombining the beams at $d$,
and observing the resulting interference fringes at $o$.  A classical
calculation shows that the motion of the earth at speed $v$ relative
to the absolute reference frame should cause the light traveling
along a path parallel to the direction of motion to take slightly
longer than light traveling along a path perpendicular to the
direction of motion, the difference in travel time being
\begin{equation}
\label{MMexpect}
\Delta t = \frac{L}{c}\cdot\frac{v^2}{c^2}
\end{equation}
to second order in $v/c$; L is the length of the interferometer
arms. By rotating the interferometer through ninety degrees,
Michelson expected to see a net shift of interference fringes equal
to 0.4 fringe.  What he measured was at most 0.01 fringe shift. This
is generally interpreted as a null result.

The analysis of this experiment is based on two implicit
assumptions: (1) there exists an absolute reference frame in which
light travels at a constant speed $c=3.00 \times 10^8 \: {\mathrm
m/s}$ \emph{relative to this reference frame}, and (2) the geometry
of the interferometer is not changed by its motion. These
assumptions play a pivotal role in the two alternative explanations
for the null result obtained in the experiment.

The first explanation was proposed by Fitzgerald\cite{Fitzgerald}
and Lorentz\cite{Lorentz}. They chose to abandon the second
assumption.  They pointed out that if the arm of the interferometer
in the direction parallel to the direction of motion shrank by a
factor of $\sqrt{1-v^2/c^2}$ that the expected fringe shift would be
exactly canceled, resulting in the observed null result.  They
interpreted this shortening of the interferometer arm as an actual
physical shortening.  Although Lorentz gave a reasonable
justification for this physical shortening in terms of his theory of
the electron, this explanation was generally considered \emph{ad
hoc} and did not gain general acceptance.

The second explanation was presented by Einstein who chose on
philosophical grounds to abandon the first assumption, simply
assuming the non-existence of an absolute reference frame. It is an
important point, however, that the special theory of relativity,
though consistent with the non-existence of an absolute reference
frame, does not prove this non-existence, only the impossibility of
detecting it from an inertial reference frame\cite{Guerra}.

The traditional analysis of the Michelson-Morley experiment assumes
that the observer and his interferometer are in an inertial
reference frame, moving at a constant speed $v$ relative to the
absolute reference frame.  The failure of the traditional analysis of
the Michelson-Morley
experiment to measure this speed, and the subsequent success of
special relativity, which raises the constancy of the speed of light
to the level of a fundamental postulate, demonstrate the
difficulty of measuring the observer's absolute speed by
performing any experiment in an inertial reference frame.

But it is well known that while constant velocity motion is
relative, accelerated motion is absolute.  Two space travelers can be
accelerating relative to each other while only one (the one with his
rockets firing) is pushed back into his seat.  The ability to detect
absolute acceleration mechanically suggests the possibility of
detecting absolute acceleration optically as well using, for
example, a Michelson-Morley interferometer.
This idea gains support from statements made by Desloge and Philpott\cite{Desloge},
\begin{quote}
In an inertial frame, distances measured with standard rods and distances made with light signals give identical results.  Furthermore, the clocks of the individual observers making up an inertial frame can be synchronized and will remain synchronized.  Neither of the above facts is true for an accelerated frame.
\end{quote}
and by Giannoni and Gr{\o}n\cite{Giannoni},
\begin{quote}
We see that these two situations, one with acceleration of the clocks and the other with the acceleration of the observer, which are identical as to the relative movements between clocks and observer, nevertheless are physically different.  In the former case, the clocks do not maintain their Einstein synchronization, while in the latter case they do so.  This illustrates that absolute acceleration has an empirical and purely kinematical significance in the special theory of relativity which is lacking in Newtonian physics.
\end{quote}

More simply, since it is well established that material objects
cannot travel faster than the speed of light, a boost given to a moving object in the forward
direction will produce a smaller change of velocity than a boost in
the backward direction. In the Michelson-Morley experiment light
travels in both the forward and backward directions and the degree
to which the acceleration of the interferometer's reference frame
affects the forward and backward travel times depends on the speed
of the interferometer. The resulting fringe shift in the
interferometer therefore depends not only on the interferometer's
acceleration but also on its absolute velocity.  This makes it
possible, in principle at least, to measure absolute velocity using
an interferometer in an accelerated reference frame.

\section{Kinematics of a Michelson-Morley interferometer in an accelerated reference frame\label{kinematics}}
In this paper we make the assumptions that (1) a preferred reference
frame exists in which the speed of light is uniform in all directions,
and (2) length contraction is a physically real phenomenon and depends
on an object's speed relative to this preferred reference frame.  The
first assumption has found support in recent work by Cahill and
Kitto\cite{Cahill}; and the second in the book by Brown\cite{Brown}
and a recent paper by Barcel\`o and Jannes\cite{Barcelo}.
We analyze a Michelson-Morley type experiment performed in a reference frame which is accelerating relative to this preferred reference frame at a rate which is constant relative to instantaneously co-moving inertial frames, otherwise known as constant proper acceleration.

Reference frames with constant acceleration relative to instantaneously co-moving inertial frames have been thoroughly investigated\cite{Moller, MTW, Hamilton, Rindler, Desloge}.  The motion of a point object with constant proper acceleration $g$ is referred to as hyperbolic motion and is described by the equation
\begin{equation}
\label{hyperg} x^2-c^2 t^2=\frac{c^4}{g^2},
\end{equation}
where the initial conditions $x=c^2/g$ and $v=0$ at time $t=0$ have been chosen.  Thus the position of this point object as a function of time is
\begin{equation}
\label{xoft} x(t)=\sqrt{\frac{c^4}{g^2}+c^2 t^2}.
\end{equation}
The velocity of this point object as a function of time is
\begin{equation}
\label{voft} v(t)=\frac{gt}{\sqrt{1+\left(\frac{gt}{c}\right)^2}},
\end{equation}
which is easily inverted to get $t$ as a function of $v$,
\begin{equation}
\label{tofv} t(v)=\frac{v}{g\sqrt{1-\left(\frac{v}{c}\right)^2}}.
\end{equation}
This last equation will be useful for explicitly showing the velocity dependence of our results.

For an extended object, such as the arm of a Michelson-Morley interferometer, the different points of the object along the direction of motion will necessarily have different accelerations because of the effects of relativistic length contraction as the object's velocity changes.  The various points along an extended object move along a family of hyperbolas described by the equation
\begin{equation}
\label{hyperX} x^2-c^2 t^2=X^2,
\end{equation}
where X is a parameter identifying the location of each point at $t=0$.

When we consider the Michelson-Morley experiment in an accelerated reference frame, we must therefore be careful to identify which point of the interferometer is moving with constant proper acceleration.  We choose the beamsplitter as this point.
If we orient the interferometer so that it is moving in the positive $x$ direction (to the right in Fig.~\ref{fig:interferometer}), with one arm parallel to the direction of motion and ahead of the beamsplitter, then the left end of the parallel arm and both ends of the perpendicular arm share the same $x$ coordinate and have motion described by the equation
\begin{equation}
\label{xlsquare} x_L^2-c^2 t^2=\frac{c^4}{g^2},
\end{equation}
or
\begin{equation}
\label{xl} x_L(t)=\sqrt{\frac{c^4}{g^2}+c^2 t^2}.
\end{equation}
The right end of the parallel arm with proper length $L$ will be located at $x=c^2/g+L$ at $t=0$.  Its motion will be described by the equation
\begin{equation}
\label{xrsquare} x_R^2-c^2 t^2=\left(\frac{c^2}{g}+L\right)^2,
\end{equation}
or
\begin{equation}
\label{xr} x_R(t)=\sqrt{\left(\frac{c^2}{g}+L\right)^2+c^2 t^2}.
\end{equation}
We assume that both arms have proper length $L$ and we can therefore identify the coordinates of the three key points of the interferometer as functions of time (as measured in the rest frame): beamsplitter at $(x_L(t),0)$, far end of perpendicular arm at $(x_L(t), L)$, and far end of the parallel arm at $(x_R(t), 0)$.

Light entering a Michelson-Morley interferometer will, in general, return to the beamsplitter at different times depending on the travel times along the two arms.  The difference between the return times can be used to calculate the phase shift in an interferometer moving with constant velocity but this method introduces a small error in the accelerated case.  When an experiment is performed, interference is observed between the two lightbeams returning to the beamsplitter at the \emph{same time}.  We take this time, which we will call $t_m$ (time of measurement), as our reference time and calculate backwards to find the difference in times of entry into the beamsplitter.  Besides giving a rigorously correct calculation of the travel time difference, this also allows us to identify unique values of the standard relativistic parameters $\beta$ and $\gamma$ as those pertaining to the time $t_m$,
\begin{equation}
\beta = \frac{v(t_m)}{c}, \qquad \gamma=(1-\beta^2)^{-\frac{1}{2}}.
\end{equation}
There are no restrictions on the value of $t_m$ and we want to consider the general case with $t_m \neq 0$ such that the interferometer has a nonzero velocity given by equation (\ref{voft}).  Furthermore, we want to consider $t_m$ as a parameter and to solve for light travel times in terms of $t_m$, thereby encoding the velocity dependence of the result.  After solving for the travel times along each arm, we will make this velocity dependence explicit by using equation (\ref{tofv}).

\section{Michelson-Morley experiment in an accelerated reference frame\label{expt}}
With the coordinates of the endpoints of the arms of a Michelson-Morley interferometer (moving with constant proper acceleration) identified as functions of time, we can now analyze the Michelson-Morley experiment in this accelerated reference frame.

We consider first light traveling along the perpendicular arm, with light entering the beamsplitter at point $a$ at time $t_1$, being reflected from the far mirror at point $b$ at time $t_3$, and returning to the beamsplitter at point $d$ at time $t_m$.  Working backward in time, light traveling from $b$ to $d$ takes time $c (t_m - t_3)$.  Time $t_3$ therefore must satisfy the equation
\begin{equation}
\label{t3x3eqn} c^2 (t_m - t_3)^2 = L^2+(x_L(t_m)-x_L(t_3))^2
\end{equation}
This can be written as an equation in $t_3$ using equation (\ref{xl}),
\begin{equation}
\label{t3eqn} c^2 (t_m - t_3)^2 = L^2+\frac{c^4}{g^2}\left[ \sqrt{1+ \left(\frac{gt_m}{c}\right)^2}-\sqrt{1+ \left(\frac{gt_3}{c}\right)^2} \right]^2.
\end{equation}
The solution of this equation for time $t_3$ is quite intricate, so we give an outline of the solution here.  We introduce the (small) dimensionless parameters
\begin{equation}
\label{dimparam} T_i = \frac{gt_i}{c} \qquad \mbox{and} \qquad \alpha = \frac{gL}{c^2},
\end{equation}
which transforms equation (\ref{t3eqn}) into a dimensionless equation,
\begin{equation}
\label{T3eqn} (T_m - T_3)^2 = \alpha^2 + \left[ \sqrt{1+T_m^2} - \sqrt{1+T_3^2} \right]^2.
\end{equation}
It is convenient to also define the quantities
\begin{equation}
\label{Xdefn} X_i = 1 + T_i^2 \qquad (i = 1, 2, 3, m),
\end{equation}
which simplifies the equation even further,
\begin{equation}
\label{T3X3eqn} (T_m - T_3)^2 = \alpha^2 + \left( X_m^{1/2}-X_3^{1/2} \right)^2.
\end{equation}
By expanding the last term, rearranging, and squaring the resulting equation, we get a polynomial equation in $T_3$ of degree 4,
\begin{equation}
\label{T3quartic} \left[ X_m + X_3 - (T_m-T_3)^2 + \alpha^2 \right]^2 - 4 X_m X_3 = 0.
\end{equation}
This quartic equation is then solved using a power series expansion of $T_3$ in powers of $\alpha$, solving for the coefficients of this power series (which are functions of $T_m$), then grouping terms into recognizable power series involving $\alpha$ and $T_m$.  The result of this is the exact solution of equation (\ref{T3quartic}),
\begin{equation}
\label{T3soln} T_3 = T_m - \alpha \sqrt{1+T_m^2} \sqrt{1+\left(\frac{\alpha}{2}\right)^2} + \frac{1}{2}T_m \alpha^2.
\end{equation}
To restore dimensioned quantities we observe that, using equation (\ref{tofv}), $\sqrt{1+T_m^2} = \gamma$ and $T_m = \beta \gamma$, and we get
\begin{equation}
\label{Tbd} t_3 = t_m - \gamma \frac{L}{c}\sqrt{1+\left(\frac{\alpha}{2}\right)^2} + \beta \gamma \frac{L}{c}\left(\frac{1}{2} \alpha \right).
\end{equation}

Light traveling from point $a$ to point $b$ takes time $c (t_3 - t_1)$.  Time $t_1$ therefore must satisfy the equation
\begin{equation}
\label{t1x1eqn} c^2 (t_3 - t_1)^2 = L^2+(x_L(t_3)-x_L(t_1))^2,
\end{equation}
which can be written as an equation in $t_1$ using equation (\ref{xl}),
\begin{equation}
\label{t1eqn} c^2 (t_3 - t_1)^2 = L^2+\frac{c^4}{g^2}\left[ \sqrt{1+ \left(\frac{gt_3}{c}\right)^2}-\sqrt{1+ \left(\frac{gt_1}{c}\right)^2} \right]^2.
\end{equation}
Following a procedure similar to the one above, we convert this to a dimensionless equation,
\begin{equation}
\label{T1X1eqn} (T_3 - T_1)^2 = \alpha^2 + \left( X_3^{1/2}-X_1^{1/2} \right)^2,
\end{equation}
expand the last term, rearrange, and square the resulting equation to get a polynomial equation in $T_1$ of degree 4,
\begin{equation}
\label{T1quartic} \left[ X_3 + X_1 - (T_3-T_1)^2 + \alpha^2 \right]^2 - 4 X_3 X_1 = 0.
\end{equation}
The exact solution of equation (\ref{T1quartic}) is
\begin{equation}
\label{T1soln} T_1 = T_3 - \alpha \sqrt{1+T_m^2} (1+\alpha^2) \sqrt{1+\left(\frac{\alpha}{2}\right)^2} + \frac{3}{2}T_m \alpha^2 + \frac{1}{2}T_m \alpha^4,
\end{equation}
which in dimensioned quantities gives
\begin{equation}
\label{Tab} t_1 = t_3 - \gamma \frac{L}{c}(1+\alpha^2) \sqrt{1+\left(\frac{\alpha}{2}\right)^2} +  \beta \gamma \frac{L}{c} \left( \frac{3}{2} \alpha + \frac{1}{2} \alpha^3 \right).
\end{equation}

The total light travel time along the perpendicular arm is $t_{perp} = t_m - t_1$:
\begin{equation}
\label{Tperp} t_{perp} = \gamma \frac{L}{c}(2+\alpha^2) \sqrt{1+\left(\frac{\alpha}{2}\right)^2} - \beta \gamma \frac{L}{c}\left(2 \alpha + \frac{1}{2} \alpha^3 \right).
\end{equation}

Now consider light traveling along the parallel arm, with light entering the beamsplitter at point $a$ at time $t_2$, being reflected from the far mirror at point $c$ at time $t_4$, and returning to the beamsplitter at point $d$ at time $t_m$.  Working backward in time, light traveling from $c$ to $d$ takes time $c (t_m - t_4)$.  Time $t_4$ therefore must satisfy the equation
\begin{equation}
\label{t4x4eqn} c (t_m - t_4) = (x_R(t_4)-x_L(t_m))
\end{equation}
This can be written as an equation in $t_4$ using equations (\ref{xl}) and (\ref{xr}),
\begin{equation}
\label{t4eqn} c (t_m - t_4) = \frac{c^2}{g}\left[ \sqrt{\left(1+\frac{gL}{c^2}\right)^2 + \left(\frac{gt_4}{c}\right)^2}-\sqrt{1+ \left(\frac{gt_m}{c}\right)^2} \right].
\end{equation}
We again transform this to a dimensionless equation,
\begin{equation}
\label{T4eqn} T_m - T_4 = \sqrt{(1+ \alpha)^2+T_4^2} - \sqrt{1+T_m^2}.
\end{equation}
We define the dimensionless parameter
\begin{equation}
\label{X4defn} X_4 = (1 + \alpha)^2 + T_4^2,
\end{equation}
which simplifies the equation even further,
\begin{equation}
\label{T4X4eqn} (T_m - T_4) = X_4^{1/2}-X_m^{1/2}.
\end{equation}
By squaring, rearranging, and squaring again, we get a polynomial equation in $T_4$ of degree 4,
\begin{equation}
\label{T4quartic} \left[ X_4 + X_m - (T_m-T_4)^2 \right]^2 - 4 X_4 X_m = 0.
\end{equation}
The exact solution of equation (\ref{T4quartic}) is,
\begin{equation}
\label{T4soln} T_4 = T_m - (\sqrt{1+T_m^2}+T_m) (\alpha + \frac{1}{2} \alpha^2).
\end{equation}
Restoring dimensioned quantities we get
\begin{equation}
\label{Tac} t_4 = t_m - \gamma \frac{L}{c}\left(1 + \frac{1}{2} \alpha \right) + \beta \gamma \frac{L}{c}\left(1 + \frac{1}{2} \alpha \right).
\end{equation}

Light traveling from point $a$ to point $c$ takes time $c (t_4 - t_2)$.  Time $t_2$ therefore must satisfy the equation
\begin{equation}
\label{t2x2eqn} c (t_4 - t_2) = x_R(t_4)-x_L(t_2)
\end{equation}
This can be written as an equation in $t_2$ using equations (\ref{xl}) and (\ref{xr}),
\begin{equation}
\label{t2eqn} c (t_4 - t_2) = \frac{c^2}{g}\left[ \sqrt{\left(1+\frac{gL}{c^2}\right)^2 + \left(\frac{gt_4}{c}\right)^2}-\sqrt{1+ \left(\frac{gt_2}{c}\right)^2} \right].
\end{equation}
Following a procedure similar to the one above, we convert this to a dimensionless equation,
\begin{equation}
\label{T2eqn} T_4 - T_2 = \sqrt{(1+ \alpha)^2+T_4^2} - \sqrt{1+T_2^2},
\end{equation}
or
\begin{equation}
\label{T2X2eqn} T_4 - T_2 = X_4^{1/2}-X_2^{1/2},
\end{equation}
square, rearrange, and square again to get a polynomial equation in $T_2$ of degree 4,
\begin{equation}
\label{T2quartic} \left[ X_4 + X_2 - (T_4-T_2)^2 \right]^2 - 4 X_4 X_2 = 0.
\end{equation}
The exact solution of equation (\ref{T4quartic}) is
\begin{equation}
\label{T2soln} T_2 = T_4 - \frac{1}{2} \left[\sqrt{1+T_m^2} + T_m \right] \left[1-\frac{1}{(1+\alpha)^2}\right].
\end{equation}
which in dimensioned quantities gives
\begin{equation}
\label{Tab} t_2 = t_4 - \gamma \frac{L}{c} \left(1 + \frac{1}{2}\alpha \right) \frac{1}{(1+\alpha)^2} - \beta \gamma \frac{L}{c} \left(1 + \frac{1}{2}\alpha \right) \frac{1}{(1+\alpha)^2}.
\end{equation}

The total light travel time along the parallel arm is $t_{par} = t_m - t_2$:
\begin{equation}
\label{Tpar} t_{par} = \gamma \frac{L}{c}\left(2 + 3\alpha + 2\alpha^2 + \frac{1}{2}\alpha^3\right) \frac{1}{(1+\alpha)^2}
- \beta \gamma \frac{L}{c}\left(2\alpha + 2\alpha^2 + \frac{1}{2}\alpha^3\right) \frac{1}{(1+\alpha)^2}.
\end{equation}

The difference in light travel times between the perpendicular arm and the parallel arm is $\Delta t = t_{perp} - t_{par} = t_2 - t_1$:
\begin{eqnarray}
\Delta t & = & \gamma \frac{L}{c}\left[(2 + \alpha^2) \sqrt{1+\left(\frac{\alpha}{2}\right)^2} - \left( 2 + 3\alpha +2\alpha^2 + \frac{1}{2}\alpha^3\right) \frac{1}{(1+\alpha)^2}\right] \nonumber \\
& & {}-\beta \gamma \frac{L}{c}\left[ 2\alpha + \frac{1}{2}\alpha^3 - \left( 2\alpha + 2\alpha^2 + \frac{1}{2}\alpha^3\right) \frac{1}{(1+\alpha)^2} \right].
\end{eqnarray}
This is an exact solution, but it obscures the dependence on the acceleration $\alpha$ because the terms that are lowest power in $\alpha$ cancel.
To see the dependence of $\Delta t$ on $\alpha$ we expand the square root and inverse square terms in power series and get
\begin{eqnarray}
\Delta t & = & \gamma \frac{L}{c} \left( \alpha - \frac{3}{4}\alpha^2 + \frac{5}{2}\alpha^3 - \frac{185}{64}\alpha^4 + \frac{7}{2}\alpha^5 - \frac{2051}{512}\alpha^6 + \ldots \right) \nonumber \\
& & \quad - \beta \gamma \frac{L}{c} \left( 2\alpha^2 - 2\alpha^3 + 3\alpha^4 - \frac{7}{2}\alpha^5 + 4\alpha^6 - \ldots \right),
\end{eqnarray}
or, keeping only the leading terms,
\begin{equation}
\Delta t = \gamma \frac{L}{c} \left( \alpha - \frac{3}{4}\alpha^2 - 2\beta \alpha^2 \right) + \mathcal{O}(\alpha^3).
\end{equation}

The analysis above has been done in the rest frame.  To determine the fringe shift observed by the experimenter in the moving reference frame we need to correct for relativistic time dilation, noting that $\Delta t = \gamma \, \Delta \tau$, where $\Delta \tau$ is the proper time interval as measured in the experimenter's moving reference frame,
\begin{equation}
\Delta \tau = \frac{L}{c} \left( \alpha - \frac{3}{4}\alpha^2 - 2\beta \alpha^2 \right) + \mathcal{O}(\alpha^3).
\end{equation}

Upon rotating the interferometer through ninety degrees (which doubles the total fringe shift) and using
light of proper wavelength $\lambda$, this produces a total observed
fringe shift of
\begin{equation}
\label{fringediffalpha}\Delta \phi = 2\frac{L}{\lambda} \left( \alpha - \frac{3}{4}\alpha^2 - 2\beta \alpha^2 \right) + \mathcal{O}(\alpha^3).
\end{equation}
We note two properties of the fringe shift given by equation
(\ref{fringediffalpha}).  First, it reduces to the observed null result
when the interferometer is moving at constant velocity ($\alpha = 0$).  Secondly,
and more importantly, the phase shift depends not only on the rate
of acceleration but also on the absolute velocity ($\beta = v/c$) of the
interferometer.

\section{Feasibility of experimental measurement of absolute velocity}
For easily accessible values of $v$ and $g$ the expected fringe
shift according to equation (\ref{fringediffalpha}) is,
unfortunately, quite small.  For a comfortable acceleration of 10 m/s$^2$ and an interferometer arm length of 10 m, the dimensionless acceleration parameter $\alpha$ is approximately $10^{-15}$; for a speed of 370 km/s (Earth's speed relative to the cosmic microwave background), the dimensionless velocity parameter $\beta$ is approximately $10^{-3}$; and for visible light the the quantity $2L/\lambda$ is approximately $3 \times 10^8$.   For these values, the three terms in equation (\ref{fringediffalpha}) are of order $10^{-8}$, $10^{-23}$, and $10^{-25}$ fringes respectively.
This makes experimental determination of
absolute velocity especially challenging, to say the least. Further complicating the problem are the mechanical stresses induced by acceleration which will cause a mechanical compression of the interferometer arm, however this can presumably be measured and compensated for.  More extreme values of acceleration, speed, and interferometer arm length could be considered, but it is difficult to imagine an experimental situation in which the velocity dependence of the fringe shift would be large enough to be detectable.

\section{Conclusions\label{conclusions}}
We conclude that it is possible in principle (though exceedingly difficult in practice) for an observer to
measure his speed relative to the absolute reference frame by
performing the Michelson-Morley experiment in an accelerated
reference frame.  This is made possible because of the coupling of
velocity and acceleration that results from the asymmetry of the
effects of acceleration in a moving reference frame.  In other
words, because there is an upper limit to achievable speeds, namely
the speed of light, it is easier to slow down than it is to speed
up.  This coupling of $v$ and $a$ is apparent in equation
(\ref{fringediffalpha}), which predicts a
non-zero velocity-dependent fringe shift in the Michelson-Morley
experiment when performed in an accelerated reference frame.

It should be emphasized that the result obtained here is not in
conflict with the \emph{limited} principle of relativity which says
that any experiment performed in a reference frame moving with
uniform velocity gives the same results as in a reference frame at
rest.  It does show, however, that this \emph{limited} principle of
relativity does not extend to accelerated reference frames.  On the
other hand, the results obtained here \emph{are} in conflict with
the broader \emph{philosophical} principle of relativity which says
that all motion is relative and that absolute motion has no physical
meaning.  It is the possibility of experimentally disproving this
broader principle which makes this result so interesting.

There is, perhaps, a more fundamental lesson to be learned here.  We have considered only one type of experiment, the Michelson-Morley experiment, and shown that, while it fails to detect absolute motion when performed in an inertial frame, it demonstrates velocity dependence when performed in an accelerated frame.  This suggests that we should go beyond the large class of experiments that have been considered in inertial frames (which all fail to detect absolute motion) and reconsider performing them in accelerated reference frames.  If we are clever enough, perhaps we can find one with enough sensitivity to measure our absolute velocity.

\section*{Acknowledgements}
I would like to thank Jackson Messner for a series of interesting
discussions on this topic.  I would also like to thank two anonymous reviewers whose suggestions have led to a significantly improved paper.

\end{document}